\documentclass{aa}
\usepackage{txfonts}
\usepackage{graphicx}
\def\I{\mbox{IRAS\,23385+6053}}

\def\Msun{\mbox{$M_\odot$}}
\def\Lsun{\mbox{$L_\odot$}}

\def\CO{$^{12}$CO}
\def\COI{\mbox{$^{13}$CO}}
\def\COII{\mbox{C$^{18}$O}}

\def\METH{CH$_3$OH}
\def\FORM{H$_2$CO}
\def\KETE{H$_2$CCO}
\def\CYAC{HC$_3$N}
\def\MCN{\mbox{CH$_3$CN}}

\def\HM{\mbox{H$_2$}}
\def\HII{H{\sc ii}}

\def\kms{\mbox{km~s$^{-1}$}}
\def\cmc{cm$^{-3}$}
\def\cmq{cm$^{-2}$}

\def\Log{\mbox{\rm Log$_{10}$}}
\def\Mdyn{\mbox{$M_{\rm dyn}$}}

\begin{document}
\title{
IRAS\,23385+6053: An embedded massive cluster in the making\thanks{Based
on observations carried out with IRAM/NOEMA. IRAM is supported
by INSU/CNRS (France), MPG (Germany), and IGN (Spain).}
}
\author{
        R.~Cesaroni\inst{1}
        \and
	H.~Beuther\inst{2}
        \and
	A.~Ahmadi\inst{2}
	\and
	M.~T.~Beltr\'an\inst{1}
        \and
	T.~Csengeri\inst{3}
        \and
	R.~Galv\'an-Madrid\inst{4}
        \and
	C.~Gieser\inst{2}
        \and
	T.~Henning\inst{2}
        \and
	K.~G.~Johnston\inst{5}
        \and
	P.~D.~Klaassen\inst{6}
        \and
	R.~Kuiper\inst{7}
        \and
	S.~Leurini\inst{8}
        \and
	H.~Linz\inst{2}
        \and
	S.~Longmore\inst{9}
        \and
	S.~L.~Lumsden\inst{5}
        \and
	L.~T.~Maud\inst{10}
        \and
	L.~Moscadelli\inst{1}
        \and
	J.~C.~Mottram\inst{2}
        \and
	A.~Palau\inst{4}
        \and
	T.~Peters\inst{11}
        \and
	R.~E.~Pudritz\inst{12}
        \and
	\'A.~S\'anchez-Monge\inst{13}
        \and
	P.~Schilke\inst{13}
        \and
	D.~Semenov\inst{2,14}
        \and
	S.~Suri\inst{2}
        \and
	J.~S.~Urquhart\inst{15}
	\and
	J.~M.~Winters\inst{16}
        \and
	Q.~Zhang\inst{17}
	\and
	H.~Zinnecker\inst{18,19}
}
\institute{
 INAF, Osservatorio Astrofisico di Arcetri, Largo E. Fermi 5, I-50125 Firenze, Italy
	   \email{cesa@arcetri.astro.it}
\and
 Max Planck Institute for Astronomy, K\"onigstuhl 17, 69117 Heidelberg, Germany
\and
 Max-Planck-Institut f\"ur Radioastronomie, Auf dem H\"ugel 69, D-53121 Bonn, Germany
\and
 Instituto de Radioastronom\'{\i}a y Astrof\'{\i}sica, Universidad Nacional Aut\'onoma de M\'exico, PO Box 3-72, 58090 Morelia, Michoac\'an, M\'exico
\and
 School of Physics and Astronomy, University of Leeds, Leeds LS2 9JT, United Kingdom
\and
 UK Astronomy Technology Centre, Royal Observatory Edinburgh, Blackford Hill, Edinburgh, EH9 3HJ, UK
\and
 Institute of Astronomy and Astrophysics, University of T\"ubingen, Auf der Morgenstelle 10, 72076, T\"ubingen, Germany
\and
 INAF, Osservatorio Astronomico di Cagliari, Via della Scienza 5, I-09047, Selargius (CA), Italy
\and
 Astrophysics Research Institute, Liverpool John Moores University, Liverpool, L3 5RF, UK
\and
 European Southern Observatory, Karl-Schwarzschild-Str. 2, D-85748 Garching bei M\"unchen, Germany
\and
 Max-Planck-Institut f\"ur Astrophysik, Karl-Schwarzschild-Str. 1, D-85741 Garching, Germany
\and
 Department of Physics and Astronomy, McMaster University, 1280 Main St. W, Hamilton, ON L8S 4M1, Canada
\and
 I. Physikalisches Institut, Universit\"at zu K\"oln, Z\"ulpicher Strasse 77, 50937, K\"oln, Germany
\and
 Department of Chemistry, Ludwig Maximilian University, Butenandtstr. 5-13, 81377 Munich, Germany
\and
 Centre for Astrophysics and Planetary Science, University of Kent, Canterbury CT2 7NH, UK
\and
 Institut de Radioastronomie Millim\'etrique (IRAM), 300 rue de la Piscine, F-38406 Saint Martin d'H\`eres, France
\and
 Center for Astrophysics $|$ Harvard \& Smithsonian, 60 Garden Street, Cambridge, MA 02138, USA
\and
 Deutsches SOFIA Institut, Pfaffenwaldring 29, Universit\"at Stuttgart, 70569 Stuttgart, Germany
\and
 Universidad Autonoma de Chile, Av. Pedro Valdivia 425, Santiago de Chile, Chile
}
\offprints{R. Cesaroni, \email{cesa@arcetri.astro.it}}
\date{Received date; accepted date}

\abstract{
This study is part of the project ``CORE'',
an IRAM/NOEMA large program consisting of observations of the millimeter
continuum and molecular line emission towards 20 selected high-mass star
forming regions. The goal of the program is to search for circumstellar
accretion disks, study the fragmentation process of molecular clumps,
and investigate the chemical composition of the gas in these regions.
}{
We focus on \I, which is believed to be the least evolved source of the
CORE sample.  This object is characterized by a compact molecular clump
that is IR dark shortward of 24~$\mu$m and is surrounded by a stellar
cluster detected in the near-IR. Our aim is to study the structure and
velocity field of the clump.
}{
The observations were performed at $\sim$1.4~mm and employed three configurations
of NOEMA and additional single-dish maps, merged with the interferometric
data to recover the extended emission. Our correlator setup covered a
number of lines from well-known hot core tracers and a few outflow tracers.
The angular ($\sim$0\farcs45--0\farcs9) and spectral (0.5~\kms) resolutions were sufficient to resolve the clump
in \I\ and investigate the existence of large-scale motions due to rotation,
infall, or expansion.
}{
We find that the clump splits into six distinct cores when observed at
sub-arcsecond resolution. These are identified through their 1.4~mm continuum
and molecular line emission. We produce maps of the velocity, line width, and
rotational temperature from the methanol and methyl cyanide lines, which allow
us to investigate the cores and reveal a
velocity and temperature gradient in the most massive core. We also find
evidence of a bipolar outflow, possibly powered by a low-mass star.
}{
We present the tentative detection of a circumstellar self-gravitating
disk lying in the most massive core and powering a large-scale outflow
previously known in the literature. In our scenario, the star powering
the flow is responsible for most of the luminosity of \I\ ($\sim$3000~\Lsun).
The other cores, albeit with masses below the corresponding virial masses,
appear to be accreting material from their molecular surroundings and are
possibly collapsing or on the verge of collapse. We conclude that we are observing a sample
of star-forming cores that is bound to turn into a cluster of massive stars.
}

\maketitle

\section{Introduction}
\label{sint}

Studies of high-mass ($\ga10^4~L_\odot$) star formation have been hindered
until recently by limited angular resolution and sensitivity. Young
early-type stars are born deeply embedded in their parental cocoons, which
makes observations longward of a few 100~$\mu$m necessary to investigate the
gas and dust emission from their natal environment. In turn, this makes
interferometric observations at (sub)mm wavelengths the only possibility
to achieve the sub-arcsecond resolutions needed to study objects typically located
at a distance of several kpc.

The advent of instruments such as the Atacama Large Millimeter/submillimeter
Array (ALMA), the NOrthern Extended Millimeter Array (NOEMA), and the
upgraded Submillimeter Array (SMA) has allowed a breakthrough in this
respect. Not only do the superior angular resolution and sensitivity make
it possible to resolve structures $\la$200~au up to distances of several kpc, but the large number of
antennas allows good sampling of the $u,\varv$ plane in a short time, enabling
surveys of a large number of targets.

\begin{figure}
\centering
\resizebox{8.5cm}{!}{\includegraphics[angle=0]{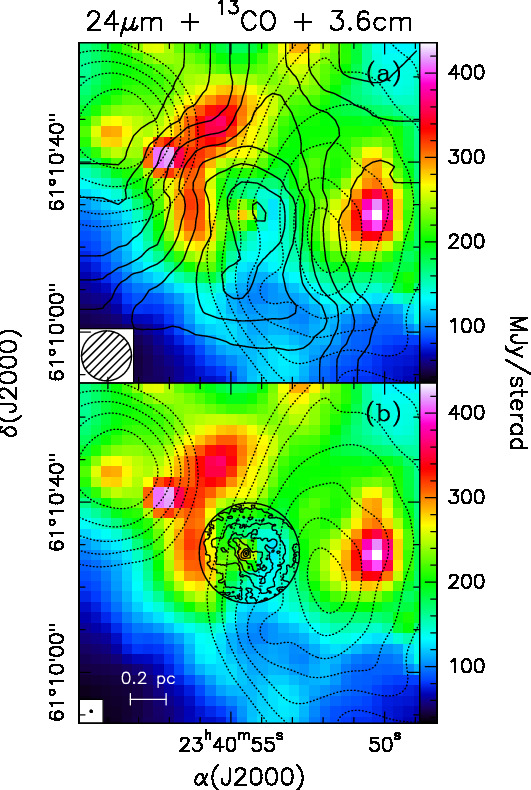}}
\caption{
{\bf a.}
Map of the emission integrated over the \COI(2--1) line (solid contours)
overlaid on an image of the continuum emission at 24~$\mu$m (from Molinari et
al.~\cite{mol08b}). The latter has an angular resolution of 6\arcsec. The
\COI\ map has been obtained with the IRAM~30-m data
a resolution of 12\arcsec\
(the beam is shown in the bottom left corner).
Solid contour levels range from 20 to 84 in steps of 8~K~\kms. The
dotted contours represent a map of the 3.6~cm continuum emission imaged
with the VLA (Molinari et al.~\cite{mol02}) and range from 0.5 to 2.3 in
steps of 0.3~mJy/beam. The synthesized beam is 10\farcs2$\times$9\farcs0
with PA=--59\degr.
{\bf b.}
Same as top panel, where solid contours are a map
of the emission averaged over the \COI(2--1) line,
obtained after merging the NOEMA with the IRAM~30-m data.
The resulting
synthesized beam is 0\farcs49$\times$0\farcs44 with PA=57\degr\
(shown in the bottom left corner).
The circle corresponds to the half-power beam width of the NOEMA antennas.
The contour levels range from 36 to 144 in steps of 18~mJy/beam.
}
\label{fbig}
\end{figure}

With this in mind, we undertook the large program ``CORE'' (P.I. Henrik
Beuther) with the IRAM interferometer, NOEMA, targeting 20 high-mass star-forming
regions. There are multiple goals of the project: to search for
circumstellar disks and study their properties; to establish the level
of fragmentation of the pc-scale molecular clumps harbouring the massive
young stellar objects (YSOs); and to understand the chemical composition
of the gas.  While some general results for the whole sample have already
been presented in a previous paper (Beuther et al.~\cite{beu18}; hereafter
BEU18), in the present study we focus on the particular source \I, in the
footsteps of other articles of the same sample (Ahmadi et al.~\cite{ahmadi2018};
Mottram et al.~submitted; Bosco et al.~submitted).

\I, located at a kinematic distance of 4.9~kpc (Molinari
et al.~\cite{mol98}), is also known as Mol160 and was selected as
a candidate massive protostar by Palla et al.~(\cite{palla91}), on
the basis of its IRAS colours. Association with water maser emission
(Palla et al.~\cite{palla91}) revealed star formation activity, later
confirmed by other studies such as Molinari et al.~(\cite{mol96}),
who detected ammonia emission from the source, indicating the presence
of dense molecular gas associated with it. Subsequently, Molinari et
al.~(\cite{mol98}) investigated the structure of the source by comparing
interferometric maps of the continuum and molecular line emission with
images of the continuum emission in the mid-IR. From this comparison one
sees that the high-mass object is embedded in an IR-dark core surrounded
by an IR-luminous stellar cluster (see Fig.~A.26 of Faustini
et al.~\cite{fau09}). While the IRAS luminosity of the region is
$\sim$$1.6\times10^4~L_\odot$, MIPSGAL/Spitzer data (Molinari et
al.~\cite{mol08b}) have demonstrated that the contribution of the core
amounts to only $\sim$$3\times10^3$~\Lsun, consistent with \I\ being a
massive protostar in the main accretion phase (Molinari et al~\cite{mol08a}).
The rest of the luminosity arises from the surrounding cluster, which
is also likely ionizing the two \HII\ regions detected by Molinari et
al.~(\cite{mol02}). Figure~\ref{fbig}a shows a pc-scale view of the region
with the molecular core, the \HII\ regions, and the far-IR emission
tracing the stellar cluster.

The presence of a compact molecular core was confirmed by the
interferometric observations at 3 and 1.3~mm of Fontani et al.~(\cite{fonta04}) and,
more recently, Wolf-Chase et al.~(\cite{wolf12}). From their observations
it was also possible to identify a main core and a secondary, less prominent
core separated by $\sim$2\arcsec\ to the NE. The temperature of the main core was
estimated to be quite low ($\sim$40~K) compared to the typical values of hot
molecular cores (100--200~K; see Kurtz et al.~\cite{ppiv}, Cesaroni~\cite{cesa05b})
and it was concluded that \I\ is indeed a rapidly accreting massive YSO in
a very early stage of its evolution. Since accretion is tightly connected
to ejection, this conclusion appears confirmed by the detection of outflow
tracers such as broad line wings and class~I methanol maser emission
associated with the cores (Molinari et al.~\cite{mol98}; Zhang et
al.~\cite{zha01,zha05}; Kurtz et al.~\cite{kur04}; Wolf-Chase et
al.~\cite{wolf12}), although a clear bipolar outflow pattern (oriented
NE--SW) is seen only on large scales ($\sim$1\farcm5) in the single-dish maps of Wu et
al.~(\cite{wu05}).
This orientation is also roughly consistent with the class~I \METH\
masers, which are distributed over a region of 6\arcsec$\times$3\arcsec
elongated in the NNE--SSW direction and centered on the molecular core
detected by Molinari et al.~(\cite{mol98}).
In summary, all these characteristics make \I\ in all likelihood the
youngest object of the CORE sample.

In the present study, we make use of the new high-resolution and high-sensitivity
NOEMA observations to pursue the investigation of \I\ and establish its
structure on scales of $\sim$0.02~pc. After describing the observations
in Sect.~\ref{sobs}, we present the results in Sect.~\ref{sres}, while
Sect.~\ref{sana} is devoted to the derivation of the physical parameters
of the cores and a detailed analysis of each of them. In Sect.~\ref{sdis}
we discuss a possible scenario for \I\ and compare its properties to those
of similar regions. Finally, a summary is given in Sect.~\ref{scon}.

\section{Observations}
\label{sobs}

\subsection{IRAM/NOEMA interferometer}
\label{spdb}

The region \I\ was observed in the framework of the NOEMA large program
CORE. Details about the program and sample can be found in BEU18. Sources
were always observed in pairs of two in the track-sharing mode. In
this case, \I\ was observed together with G108.7575--0.986 in three
configurations (A, C and D) between January 2015 and October 2016. 
The number of antennas in the array varied between 6 and 8.  The phase
center for \I\ was $\alpha$(J2000.0) 23$^{\rm h}$40$^{\rm m}$54\fs400 
and $\delta$(J2000.0) +61\degr10\arcmin28\farcs020 with the velocity of
rest of --50.2~\kms. As gain calibrator, we used the quasar 0059+581,
and bandpass and flux calibrators were 3C\,454.3 and MWC\,349.

The spectral range covered with the broad-band WIDEX correlator was between
217.167 and 220.834\,GHz at 1.95\,MHz spectral resolution (corresponding
to $\sim$2.7~\kms). Furthermore, eight narrow-band high-spectral
resolution units ($\sim$0.5~\kms) were distributed over the bandpass,
mainly covering lines from CH$_3$CN and H$_2$CO (for more details see
Table~2 in BEU18). 

Data calibration and imaging were mainly conducted within the GILDAS
framework, with the calibration package CLIC, and the imaging program
MAPPING. The continuum was produced by collapsing the line-free part of
the spectrum into a single continuum channel. Furthermore, the continuum
data were self-calibrated in CASA to improve the signal-to-noise ratio. The
continuum $1\sigma$ rms of the final images is 0.11~mJy~beam$^{-1}$, and
for the line cubes we achieved a typical rms of $\sim$6~mJy~beam$^{-1}$
in a 0.5~\kms\ channel. The spatial resolutions of the continuum and line
data are 0\farcs46 and $\sim$0\farcs9, respectively, due to the different
weighting adopted (uniform for the continuum, robust=5 for the lines).
The absolute flux scale is estimated to be correct to within 20\%.

\subsection{IRAM/30-m telescope}
\label{spico}

The above interferometer data were complemented with short spacing
observations at the IRAM 30-m telescope. Details about the 30-m
observations can be found in Ahmadi et al.~(\cite{ahmadi2018}) and
Mottram et al. (submitted). Each source was observed in the 1.4~mm
band in the On-The-Fly (OTF) observation mode in map sizes of typically
1\arcmin$\times$1\arcmin. OTF maps were conducted in both right ascension
and declination directions to reduce scanning effects. Calibration of the
single-dish data was done in CLASS, and the merging of the single-dish and
interferometer data was again conducted within the MAPPING program. While
most compact structures were still cleaned with the CLARK-algorithm,
the more extended \COI(2--1) data were imaged with the Steer-Dewdney-Ito
(SDI) method.  Details about the merging and imaging are given in Mottram
et al. (submitted). The resulting angular resolution and $1\sigma$ rms in
a 3~\kms\ channel are $\sim$0\farcs46 and 2~mJy~beam$^{-1}$, respectively.

\begin{figure} 
\centering
\resizebox{8.5cm}{!}{\includegraphics[angle=0]{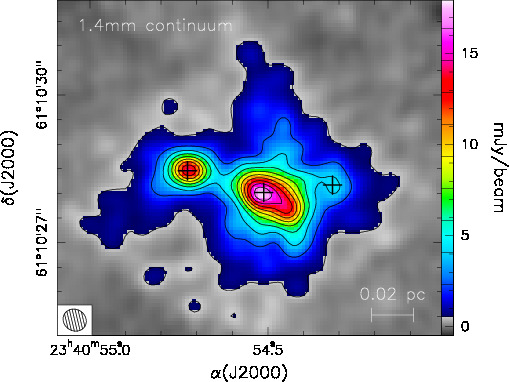}}
\caption{
Map of the 1.4~mm continuum emission from \I. The values of the contour
levels are marked in the colour scale to the right. The minimum contour
level (0.55~mJy/beam) corresponds to 5$\sigma$. The crosses indicate the
positions of the 3 cores identified by BEU18, corresponding to A1, B and E
in Table~\ref{tcores}.
}
\label{fcont}
\end{figure}

\section{Results}
\label{sres}

Below we present our main findings. Unless otherwise specified, the maps
used in our study are those obtained with the NOEMA interferometer, because
only the emission of the CO isotopologues may require merging the NOEMA
and 30-m data to be properly imaged.

\subsection{Continuum emission}
\label{scont}

A study of the continuum emission from \I\ and the other sources of
the CORE sample has already been presented by BEU18. Using the {\sc clumpfind}
algorithm, they identified 3 cores, whose total flux density is
$\sim$180~mJy. This is greater than the value of $\sim$150~mJy obtained by
Fontani et al.~(\cite{fonta04}). Such a difference is larger than the noise
of Fontani et al.'s map ($\sim$2~mJy/beam) and can be due to the different
$u,\varv$ coverage of the two data sets. In fact at the time of Fontani
et al.'s observations the Plateau de Bure interferometer was equipped with
only 5 antennas and 2 configurations of the array were used, compared with
our NOEMA observations performed with a number of antennas ranging from 6
to 8 and 3 different configurations. In particular, unlike Fontani et al.,
we used also the most compact (D-array) configuration and it is thus not
surprising that part of the flux density present in our images is filtered
out in their data.  We show in Fig.~\ref{fcont} the map of the continuum
emission, with overlaid the positions of the three cores identified by BEU18.

\begin{figure*}
\centering
\resizebox{18cm}{!}{\includegraphics[angle=0]{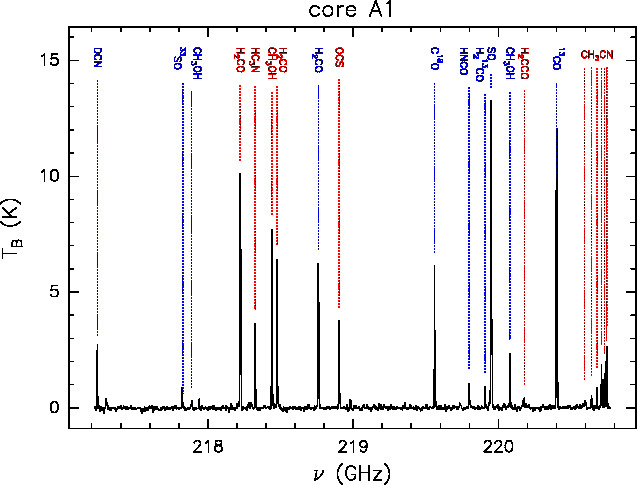}}
\caption{
Spectrum obtained with the WIDEX correlator at low spectral resolution towards the peak
of core A1. Only the strongest molecular transitions are indicated. The red
colour denotes the lines observed also at high spectral resolution.
}
\label{fspta1}
\end{figure*}

\subsection{Line emission}
\label{sline}

While the correlator setup covers many potentially useful molecular lines,
only some of these have been detected with a typical 1$\sigma$ noise of
3~mJy/beam ($\sim$0.1~K) in a 3~\kms\ channel. The most prominent are the
\COI\ and \COII\ (2--1) transitions. However, these CO isotopologues are
largely sensitive to extended emission, which masks the compact emission from
the core. In Fig.~\ref{fbig} we compare a map of the mean emission in the
\COI(2--1) line to the Spitzer/MIPS image of the region at 24~$\mu$m. More
precisely, in Fig.~\ref{fbig}a we show the \COI\ map made with the 30-m
telescope, while in Fig.~\ref{fbig}b the same map has been obtained after
merging the 30-m and NOEMA data. One can see that the compact core is
embedded in a larger emitting region extending over at least $\sim$20\arcsec\
(0.48~pc). As noted by Molinari et al~(\cite{mol08b}), the core is detected
at 24~$\mu$m (whereas it is IR-dark at shorter wavelengths). It is worth
noting that no clear bipolar outflow structure can be identified from our
CO data,
neither in the channel maps nor in the spectra.
Although this might be due to the limited spectral resolution
(3~\kms) and the relatively small region covered by the observations,
one must also consider that \CO\ would be a better outflow tracer
than its isotopologues and the outflow orientation could also prevent
detection of it, if the axis lies close to the plane of the sky and/or
the gas is ejected at low velocity.
The \COI\ emitting region appears to lie between the two compact \HII\
regions (dotted contours in Fig.~\ref{fbig}) imaged at 3.6~cm by Molinari
et al.~(\cite{mol02}), which are ionized by B-type stars belonging to the
surrounding cluster.

\begin{figure*}
\centering
\resizebox{16cm}{!}{\includegraphics[angle=0]{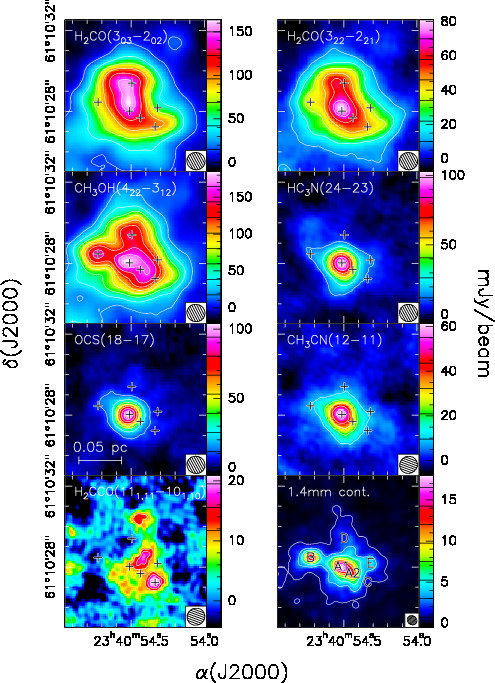}}
\caption{
Maps of the mean emission in various molecular lines. For the sake of direct
comparison with the line maps we show again the map of the continuum emission in the
bottom right panel. For each map, the values of the contour levels are marked in the
corresponding colour scale. The crosses and letters indicate the positions
of the 6 cores identified by us (see also Fig.~\ref{fske}). Cores A1, B, and E correspond
to the crosses in Fig.~\ref{fcont}. The ellipse in the
bottom right of each panel is the synthesized beam.
}
\label{flines}
\end{figure*}

The present study is focused on compact cores. These are much better traced
by other molecular species than the CO isotopologues. Our correlator
setup covers several transitions of different molecular species, as one
can see in the broad-band spectrum shown in Fig.~\ref{fspta1}, obtained
with the WIDEX correlator towards the peak of the emission. However,
for our purposes we will focus mostly on the lines covered also by the
high spectral resolution (0.5~\kms) units of the narrow-band correlator and marked in
red in Fig.~\ref{fspta1}. These belong to formaldehyde (\FORM), methanol
(\METH), methyl cyanide (\MCN), cyanoacetylene (\CYAC), carbonyl sulfide
(OCS), and ketene (\KETE). The maps of the emission averaged over the
lines of these species are shown in Fig.~\ref{flines}. Visual inspection
of these images and the continuum emission map in Fig.~\ref{fcont}, allows
us to identify 6 cores, whose positions are marked in
Fig.~\ref{flines}
with crosses.

More in detail, cores A1+A2, B, and E have been found by BEU18 from the
continuum map and are clearly seen in the \METH\ map in Fig.~\ref{flines},
core D is well defined in the \METH\ map, and core C appears as a separate
entity only in the \KETE\ map. We
point out
that we have decided to split core~1
in Table~A.1 of BEU18 into two cores, A1 and A2, because the head-tail
structure seen in the continuum is not reproduced in some of the lines. In
particular, the \CYAC, OCS, and \MCN\ emission seems to trace a circular,
barely resolved core located at the position of the continuum peak (A1).

\begin{figure}
\centering
\resizebox{8.5cm}{!}{\includegraphics[angle=0]{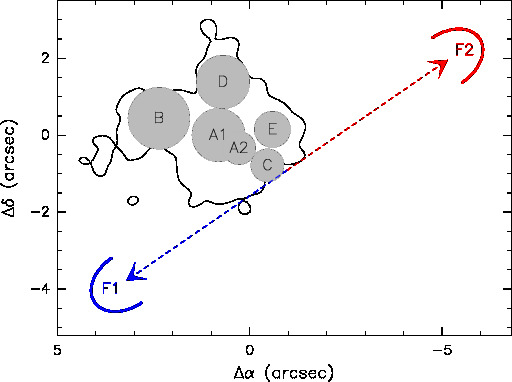}}
\caption{
Sketch of the cores identified by us in \I. The contour corresponds to
the 5$\sigma$ level of the continuum emission.
The two arrows and arcs indicate the direction and location of the expanding
lobes (F1 and F2) of the bipolar outflow detected in the \FORM\ and \METH\
emission, perhaps originating from core C.
}
\label{fske}
\end{figure}

Not all molecules are detected in all cores. Table~\ref{tcores} lists
the tracers revealed in each core, while Fig.~\ref{fske} illustrates the
locations of the cores through a sketch, to facilitate
core identification and comparison with the maps.

\begin{table*}
\caption[]{
List of tracers detected in each core and corresponding parameters. The question marks indicate
uncertain (non)detections, mostly due to the lack of a clear peak
of emission coinciding with the core.
}
\label{tcores}
\begin{tabular}{lccccccccc}
\hline
\hline
 tracer  & $\nu$~(MHz) & $E_{\rm up}$~(K) & $S\,\mu^2$~(D$^2$) & A1 & A2 & B & C & D & E \\
\hline
 \FORM($3_{03}$--$2_{02}$) & 218222.192 &  20.96 & 16.3 & Y  & Y  & Y  & Y  & Y  & Y? \\
 \FORM($3_{22}$--$2_{21}$) & 218475.632 &  68.09 & 9.06 & Y  & Y  & Y  & Y  & Y  & Y? \\
 \METH($4_2$--$3_1$)~E1    & 218440.050 &  34.98 & 3.48 & Y  & Y  & Y  & Y  & Y  & Y? \\
 \CYAC(24--23)             & 218324.723 & 130.98 & 334 & Y  & Y? & Y  & N  & Y  & Y? \\
 OCS(18--17)               & 218903.356 &  99.81 & 9.21 & Y  & N? & N? & N  & N  & N \\
 \MCN(12--11)              & 220747.261$^a$ & 68.87$^a$ & 254$^a$ & Y  & Y? & N? & N  & N  & N? \\
 \KETE($11_{1,11}$--$10_{1,10}$) & 220177.569 & 76.46 & 66.0 & Y  & N? & N  & Y  & N  & N \\
 1.4~mm continuum          & 218917     & --- & --- & Y  & Y  & Y  & N? & Y? & Y \\
\hline
\hline
\end{tabular}

\vspace*{1mm}
$^a$~parameters of the $K$=0 component, but emission was detected up to the $K$=6 transition
\end{table*}

An interesting feature that can be seen in the maps of the \FORM\ and \METH\
emission is the presence of two blobs, F1 and F2 (see Fig.~\ref{foutf}),
separated by $\sim$10\arcsec\ or 0.24~pc and symmetrically disposed to
the SE and NW of core C. This is shown in Fig.~\ref{foutf}, where the
dashed line joins the two blobs. While these might be two additional
cores, the non detection of continuum emission from them hints at a
different nature. We propose that they could be the tips of the lobes
of a bipolar outflow originating from C or a nearby, low-mass undetected
core. Indeed, the sensitivity (at a 5$\sigma$ level) of our continuum map
is $\sim$0.1--0.5~\Msun\ for a dust temperature in the range 20--70~K. These
values are typically above the mass of envelopes/disks around class~I YSOs.
We will further discuss
this outflow hypothesis in Sect.~\ref{sana}, whereas in Sect.~\ref{sao}
we will comment on the possible existence of another outflow powered by
a YSO in core~A1.

\section{Analysis}
\label{sana}

The goal of our study is to establish the nature of the cores identified
in our maps. In the following we derive the core physical parameters and
analyse each of them in more detail.

\subsection{Physical parameters of the cores}
\label{scpar}

The observed ratio between the brightness temperatures of the \FORM\
($3_{22}$--$2_{21}$) and ($3_{03}$--$2_{02}$) transitions over the cores
is in all cases $>$0.56, the maximum LTE ratio expected in the optically thin
case\footnote{
If the emission is thin, the ratio between the brightness temperatures of two
lines is equal to the ratio between the corresponding optical depths, therefore:
$T_{\rm B}^2/T_{\rm B}^1\simeq [(S\mu^2)_2/(S\mu^2)_1)] \, (\nu_2/\nu_1) \,
\exp[-(E_2-E_1)/T] < [(S\mu^2)_2/(S\mu^2)_1)] \, (\nu_2/\nu_1) \simeq 0.56$,
where the line strengths and frequencies are given in Table~\ref{tcores}.
}.
This indicates that the lines must be optically thick and their ratio
cannot be used to estimate the gas temperature and column density of the
cores. For the same reason, \FORM\ is not suitable for the study of the
velocity field, because it traces the envelope around the cores as suggested
by the size of the \FORM\ emission in Fig.~\ref{flines}. From this figure,
one sees that most of the cores are recognizable in the methanol map and
we thus prefer to use this species to obtain a picture of the core physical
properties. In Fig.~\ref{fmoms} we show maps of the line velocity and full
width at half maximum obtained from the 1st and 2nd moments of the \METH\
line. Overlaid on this is the map of the zero moment (integrated intensity)
in the same line.

\begin{figure}
\centering
\resizebox{8.5cm}{!}{\includegraphics[angle=0]{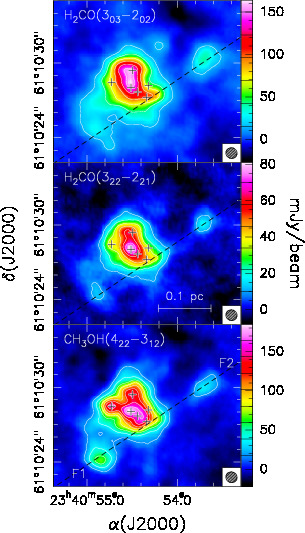}}
\caption{
Same as Fig.~\ref{flines} but over a larger region and only for the \FORM\
and \METH\ lines. The dashed line denotes the direction defined by the
two emission blobs, F1 and F2, located respectively to the SE and NW of
the main cores.
}
\label{foutf}
\end{figure}

\begin{figure}
\centering
\resizebox{8.5cm}{!}{\includegraphics[angle=0]{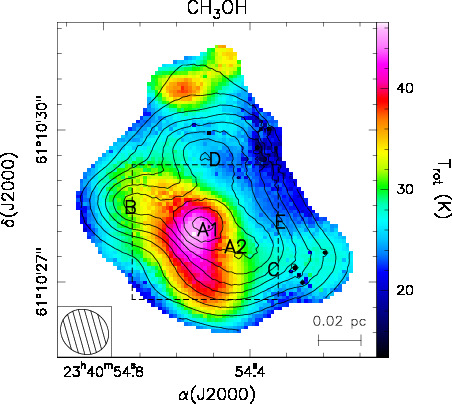}}
\caption{
Maps of the \METH\ rotational temperature (colour image) and column density
(contours) obtained with the XCLASS program. The labels indicate the cores.
Contour levels range from
$2\times10^{15}$ to $1.1\times10^{16}$ in steps of $10^{15}$~cm$^{-2}$.
Typical uncertainties on the values of the rotational temperature
are 10--20\%. The ellipse in the bottom left denotes the synthesized beam.
The dashed box corresponds to the region shown in Fig.~\ref{ftmcn}.
}
\label{ftmeth}
\end{figure}

\begin{figure}
\centering
\resizebox{8.5cm}{!}{\includegraphics[angle=0]{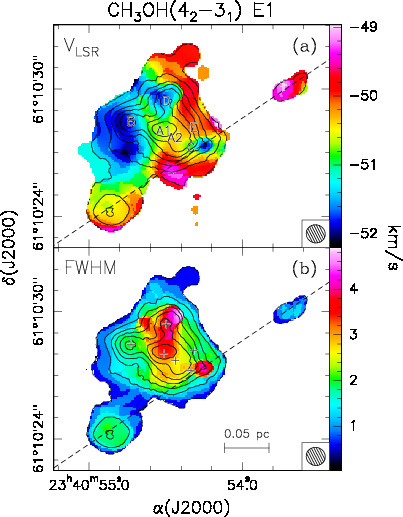}}
\caption{
{\bf a.} Map of the zero moment (integrated intensity; contours) of the \METH($4_2$--$3_1$)~E1 line
overlaid on the map (colour image) of the first moment of the same transition.
Contour levels range from 0.05 to 1.25 in steps of 0.2~Jy~beam$^{-1}$~\kms. The ellipse
in the bottom right represents the synthesized beam. The dashed line
has the same meaning as in Fig.~\ref{foutf}.
{\bf b.} Same as top panel, for the second moment of the methanol line.
}
\label{fmoms}
\end{figure}

As one can see from Fig.~\ref{fspta1}, two more lines of methanol are
detected in the bandwidth covered by the WIDEX correlator: the $8_0$--$7_1$~E2 and
$20_1$--$20_0$~E1 transitions. One can use
these transitions to calculate the rotational temperature, $T_{\rm rot}$,
of the \METH\ gas all over the cores. For this purpose we have used the
eXtended CASA Line Analysis Software Suite (XCLASS) tool\footnote{
XCLASS is available at https://xclass.astro.uni-koeln.de
}
(M\"oller et
al.~\cite{xclass}), which simultaneously fits the lines of a molecular
species assuming LTE
by
varying the relevant physical parameters, namely the
source angular size, rotational temperature, column density of the molecule,
systemic LSR velocity, and line width. One of the advantages with respect to
rotation diagrams is that XCLASS takes into account the line opacities in the
fit. In our case, the region over which \METH\ is detected is much greater
than the synthesized beam and we have thus assumed a beam filling factor
equal to 1. The maps of $T_{\rm rot}$ and \METH\ column density are shown
in Fig.~\ref{ftmeth}, where both quantities peak towards the position of A1.

\begin{figure}
\centering
\resizebox{8.5cm}{!}{\includegraphics[angle=0]{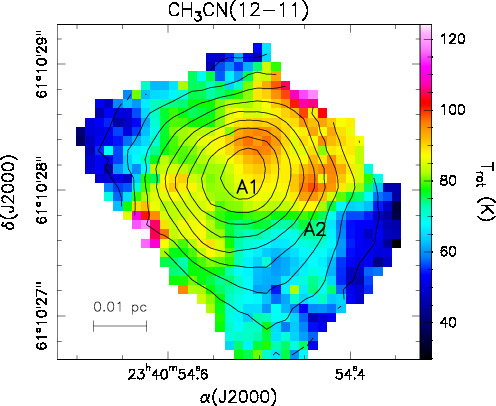}}
\caption{
Maps of the \MCN\ rotational temperature (colour image) and column density
(contours) obtained with the XCLASS program. The labels
have the same meaning as in Fig.~\ref{flines}.
Contour levels range from
$2\times10^{13}$ to $10^{14}$ in steps of $10^{13}$~cm$^{-2}$.
Typical uncertainties on the values of the rotational temperature
are 10--20\%.
}
\label{ftmcn}
\end{figure}

The same method can be applied to the \MCN(12--11) $K$=0--6 transitions, which
are known to be excellent hot core tracers. As such, these are well
suited to estimate the physical parameters of chemically rich cores
like A1 and A2 and indeed the \MCN\ emission is detected only towards
these two cores. We present the corresponding $T_{\rm rot}$ and column
density maps in Fig.~\ref{ftmcn}. The obvious difference with respect to
Fig.~\ref{ftmeth} is that the temperature obtained from \MCN\ is about twice
as much as that derived from \METH. While this could be due to methanol
being sub-thermally excited (Bachiller et al.~\cite{bachi95}; Kalenskii \&
Kurtz~\cite{kaku16}), it is most likely that the interplay between opacity
and temperature gradients plays a dominant role. Unlike the other cores,
A1 coincides with the peak of column density and it is thus possible
that species with different abundances (and opacities) trace regions with
different temperatures.

Now we compute the mean values of the physical parameters of
the cores. For this purpose, it is necessary to establish the border
of each core. In order to simplify the problem, we assume the cores
to be spherical, so that all we need to estimate is the core radius.
In Table~\ref{tparms} we give the peak positions of the cores determined
from the line and continuum maps. In particular, the positions of cores
A1, B, and E are taken from Table~A.1 of BEU18. These authors used {\sc
clumpfind} to determine the core sizes, but this method was applied to
the continuum image where only 3 cores were identified. Here we prefer to
use a different approach. For each core, we consider the separation from
the nearest of the other cores and assume that this is twice the radius of
it. We give the radii, $R_{\rm c}$, in Table~\ref{tparms}. These values
should be considered with two caveats in mind. On the one hand, the
method used by us overestimates the radius, because two neighbourhing
cores are not necessarily in contact. On the other hand, we underestimate
the separation because we see only the projection of it on the plane of the sky.

We calculate the mean values of the LSR velocity ($V_{\rm LSR}$), line
full width at half maximum ($\Delta V$), and temperature ($T_{\rm rot}$)
by averaging the corresponding \METH\ parameters (see Figs.~\ref{ftmeth}
and~\ref{fmoms}) over the surface of each core. Finally, we obtain
the total core flux densities ($S_\nu$) by integrating the continuum
emission (Fig.~\ref{fcont}) inside the borders of the cores. All these
quantities are listed in Table~\ref{tparms}, where we give also virial
masses, $M_{\rm vir}$, and the core masses, $M_{\rm H_2}$, computed from
the temperature, line widths, and flux densities according to Eq.~(3) of
MacLaren et al.~(\cite{mclar}) and Eq.~(1) of Schuller et al.~\cite{schul09},
under the assumptions of constant density, gas-to-dust mass ratio of 150,
and dust opacity of 0.9~cm$^2$~g$^{-1}$ at 1.4~mm. In the same table we
report the \HM\ densities, for a mean molecular weight of 2.8, and the
ratios $M_{\rm H_2}/M_{\rm vir}$.

\begin{table*}
\caption[]{
Parameters of the cores in the \I\ region.
}
\label{tparms}
\begin{tabular}{lccccccccccccc}
\hline
\hline
 core  & $\Delta\alpha$$^a$ & $\Delta\delta$$^a$ & $R_{\rm c}$ & $V_{\rm LSR}$$^b$ & $\Delta V$$^b$ & $T_{\rm rot}$ & $S_\nu$ &
 $M_{\rm vir}$$^c$ & $M_{\rm H_2}$ & $n_{\rm H_2}$ & $\frac{M_{\rm H_2}}{M_{\rm vir}}$ & $\dot{M}_{\rm acc}$ & $t_{\rm acc}$ \\
       & (\arcsec) & (\arcsec) & (\arcsec) & (\kms) & (\kms) & (K) & (mJy) &
 (\Msun) & (\Msun) & (\cmc) &  & (\Msun~yr$^{-1}$) & (yr) \\
\hline
 A1 &   0.81 &   0.00 & 0.69 & --50.6 & 3.2 & 39 & 62 & 35 & 24 & $1.8\times10^7$ & 0.67 & $3.1\times10^{-3}$ & $7.6\times10^3$ \\
 A2 &   0.28 & --0.34 & 0.43 & --50.4 & 2.8 & 33 & 25 & 17 & 12 & $3.8\times10^7$ & 0.70 & $2.1\times10^{-3}$ & $5.7\times10^3$ \\
 B  &   2.36 &   0.44 & 0.81 & --51.6 & 1.9 & 31 & 36 & 15 & 18 & $9.0\times10^6$ & 1.25 & $6.5\times10^{-4}$ & $2.8\times10^4$ \\
 C  & --0.46 & --0.79 & 0.43 & --50.8 & 2.7 & 27 & 3.8 & 16 & 2.2 & $7.3\times10^6$ & 0.14 & $1.9\times10^{-3}$ & $1.2\times10^3$ \\
 D  &   0.69 &   1.38 & 0.69 & --51.3 & 3.6 & 26 & 12 & 45 & 7.4 & $5.7\times10^6$ & 0.17 & $4.4\times10^{-3}$ & $1.7\times10^3$ \\
 E  & --0.59 &   0.15 & 0.47 & --50.2 & 2.0 & 24 & 10 & 9.4 & 6.8 & $1.7\times10^7$ & 0.72 & $7.6\times10^{-4}$ & $9.0\times10^3$ \\
\hline
\hline
\end{tabular}

\vspace*{1mm}
$^a$~offsets are relative to the phase center $\alpha$(J2000.0)=23$^{\rm h}$40$^{\rm m}$54\fs400 
$\delta$(J2000.0)=+61\degr10\arcmin28\farcs020 \\
$^b$~obtained from \METH \\
$^c$~computed from Eq.~(3) of MacLaren et al.~(\cite{mclar}),
assuming constant density and using the line width of \METH
\end{table*}

Clearly, the velocity changes significantly from core to core, with cores A1,
A2, and E being red-shifted by at least $\sim$1~\kms\ with respect to B and
D, while cores A1 and D are those with the largest line width and possibly
the highest level of turbulence. We note that in all cases the line is wider than
expected from pure thermal broadening ($<$0.42~\kms\ for $T<120$~K -- see
Fig.~\ref{ftmcn}), which implies a contribution from non-thermal motions. The
line width is also large close to core~C (see Fig.~\ref{fmoms}b), where
a blue-shifted ``spot'' is seen in the velocity map (Fig.~\ref{fmoms}a;
see also Sect.~\ref{scc}). Whether core~C is really associated with such
a feature is questionable. We remind the reader that this core has been
identified only from the faint \KETE\ emission, which makes it difficult to
determine a precise position for it. In any case, it appears that close to
the geometrical center of the (putative) outflow there is a small region
with a large velocity dispersion, which could contain the source powering
the flow. An alternative possibility is that this ``spot'' is where the
flow from C impinges on the dense gas, thus causing the observed enhancement of
the line width. It is worth noting that Wolf-Chase et al.~(\cite{wolf12})
have detected an H$_2$ knot (MHO~2921 in their notation) close to core C,
which is probably due to emission from shocks.

The velocity difference between blobs F1 and F2 is small,
only $\sim$1~\kms, but this is expected if the outflow lobes lie close
to the plane of the sky. However, very little velocity dispersion (the
line FWHM is $\sim$2~\kms) is observed towards these blobs, in contrast
with the expected line broadening at the tips of a bipolar flow. We will
further discuss the outflow hypothesis in Sect.~\ref{scc}.

\begin{figure}
\centering
\resizebox{8.5cm}{!}{\includegraphics[angle=0]{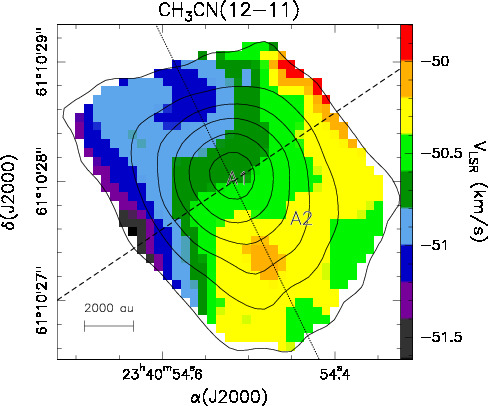}}
\caption{
Map of the \MCN(12-11) line emission averaged over the $K$=0 and 1
components (contours) overlaid on the map of the velocity in the same
line, obtained by simultaneously fitting the $K$=0 to 4 components as
explained in the text. Labels A1 and A2 mark the center positions of the
corresponding cores. Contour levels range from 9.5 (5$\sigma$) to 57 in
steps of 9.5~mJy~beam$^{-1}$. The dotted and dashed lines indicate the approximate
directions of the two velocity gradients.
}
\label{fvmcn}
\end{figure}

\subsection{Core A1}
\label{sao}

As already explained, we prefer to identify two cores (A1 and A2) where the analysis
of BEU18 finds only one. This choice, based on the
morphology of the line emission, is further confirmed by the different
kinematical properties, recognizable in Fig.~\ref{fmoms} especially for
the line FWHM which is larger (by $\sim$1~\kms) in A1, compared to A2.

The velocity field in A1 can be better analysed by means of high
density/temperature tracers such as \MCN. In Fig.~\ref{fvmcn} we show a
map of the peak velocity of the \MCN(12--11) transition obtained by fitting
Gaussian profiles simultaneously to the strongest ($K=0$ to 4) components,
after fixing the line separations to the laboratory values and forcing
the line widths to be identical. This method has the advantage of reducing
the velocity and line width uncertainties with respect to fitting each $K$
component independently or computing the first and second moments of the
lines. The velocity pattern presents two red-shifted and two blue-shifted
peaks defining two velocity gradients along directions (see the dotted and
dashed lines in Fig.~\ref{fvmcn}) roughly perpendicular to each other. The
\MCN\ emission (see contour map in Fig.~\ref{fvmcn}) clearly peaks towards
A1, which proves that this emission is largely dominated by A1 and the \MCN\
velocity is affected only marginally by the gas in A2.
In the following, we discuss three scenarios that could explain the
observed velocity field in core A1.

\subsubsection{Keplerian disk with expansion}
\label{skepexp}

The presence of two velocity gradients, directed SE--NW and SW--NE, could
be due to the combination of Keplerian rotation and expansion along the
surface of the disk, a situation reminiscent of disk winds. 

It is worth noting that we cannot distinguish between expansion and infall
on the basis of the observed velocity pattern, since in both cases the
projected velocities along the line of sight give origin to the same type
of velocity field. In practice, reversing the sign of the radial velocity
from positive (for expansion) to negative (for infall) only swaps the red-
and blue-shifted sides of the disk, which can be compensated with a rotation
by 180\degr\ of the position angle of the disk. Here, we rule out infall a
priori, because the existence of a pair of blue- and red-shifted velocity
peaks close to the disk border can be explained only if the radial velocity
component increases with radius. This situation is in contrast with
infall, which accelerates towards the star, whereas it is acceptable
for expansion, where the material is accelerated outward.

\begin{figure}
\centering
\resizebox{8.5cm}{!}{\includegraphics[angle=0]{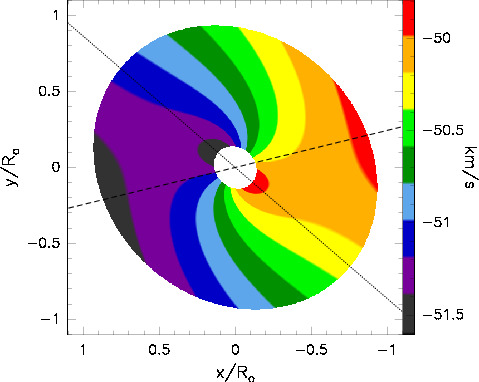}}
\caption{
Model velocity map for a geometrically thin disk undergoing both
Keplerian rotation and expansion with radial velocity proportional to the
radius. Fiducial values have been assumed for the parameters (see text).
The map has been rotated by 45\degr\ to ease the comparison with Fig.~\ref{fvmcn}.
The dotted and dashed lines indicate the directions of the two velocity
gradients defined by the blue- and red-shifted velocity peaks.
}
\label{fmod}
\end{figure}

While a physical model
to reproduce the observed velocity and intensity goes beyond the scope
of the present article, we can compare the velocity map in Fig.~\ref{fvmcn}
with the map of the line-of-sight velocity computed from a purely kinematical
model assuming Keplerian rotation and radial expansion in a geometrically
thin disk.
The observed velocity is given by the expression
\begin{equation}
V(x,y) = V_{\rm sys} + \varv_{\rm rot} \left(\frac{R}{R_{\rm o}}\right)^\alpha \frac{x}{R} \sin\theta
       + \varv_{\rm rad} \left(\frac{R}{R_{\rm o}}\right)^\beta \frac{y}{R} \tan\theta
\end{equation}
where $x$ and $y$ are cartesian coordinates in the plane of the sky
with $y$ along the the projected major axis of the disk, $V_{\rm sys}$
is the systemic velocity with respect to the local standard of rest
(LSR), $R=\sqrt{x^2+(y/\cos\theta)^2}$, $R_{\rm o}$ is the outer radius
of the disk, $\varv_{\rm rot}$ and $\varv_{\rm rad}$ are the azimuthal
and radial velocity components in the plane of the disk at $R_{\rm o}$,
$\theta$ is the angle between the disk axis and the line of sight, and
we have assumed $\alpha=-1/2$ for Keplerian rotation and $\beta=1$ for
expansion. In Fig.~\ref{fmod} we show the velocity map obtained for fiducial
values of the parameters ($\theta=30\degr$, $\varv_{\rm rot}=0.7$~\kms,
$\varv_{\rm rad}=2$~\kms, $V_{\rm sys}=-50.7$~\kms), obtained by visual
comparison with Fig.~\ref{fvmcn}. In particular, the inclination $\theta$
has been obtained from the ratio between the minor axis ($\sim$2\arcsec) and
the major axis ($\sim$2\farcs3) of the velocity map in Fig.~\ref{fvmcn},
as $\theta=\arccos(2/2.3)$. Note that all lengths are normalized with
respect to $R_{\rm o}$ and the velocity is computed down to a minimum
radius $R_{\rm i}=0.15\,R_{\rm o}$.

The model presents a number of features which are clearly not in agreement
with the data. For example, the two inner velocity peaks are too close to
the center of the disk and the directions of the two velocity gradients
(dashed and dotted lines in Fig.~\ref{fmod}) are not perpendicular to each
other as they are in Fig.~\ref{fvmcn}. However, we stress that
the comparison between Fig.~\ref{fvmcn} and Fig.~\ref{fmod} is only
qualitative, as our calculations do not take into account the angular
resolution of the observations and no attempt is made to compute the line
intensity and width. We do not want to fit the data but only to show that
rotation plus expansion could mimic the existence of two pairs of blue-
and red-shifted velocity peaks (connected by the dotted and dashed lines in
Fig.~\ref{fmod}), one located inside the disk, the other close to the
border of it. These double peaks generate the twist in the velocity
pattern that is seen in the data.

Since in this model the disk is undergoing Keplerian rotation, one can
compute the stellar mass from the high-velocity peaks close to the center of
the disk. From Fig.~\ref{fvmcn} one obtains a rotation velocity\footnote{
The velocity is obtained as half the difference between the velocities
observed at the blue-shifted ($\sim$--51~\kms) and red-shifted
($\sim$--50~\kms) peaks located along the dotted line in Fig.~\ref{fvmcn}.
}
of 0.5~\kms\ at a radius of 0\farcs73 (or 3600~au), which implies a dynamical mass
$\Mdyn\sim$1~\Msun. This value appears by far too small compared to the mass
of the core ($\sim$24~\Msun; see Table~\ref{tparms}), to satisfy the condition $M_{\rm
star}>M_{\rm disk}$ for Keplerian rotation. Moreover, it is very unlikely that
a solar type star associated with a chemically rich molecular core (``hot
corino'') could be detected at a distance of 4.9~kpc. Finally, the luminosity
of the source (3000~\Lsun) is 3 orders of magnitude greater than that of
a solar-type star.

Most of these problems can be solved if the disk is sufficiently inclined with
respect to the line
of sight. A luminosity of 3000~\Lsun\ star corresponds to a stellar mass
$M_\ast\simeq9$~\Msun\ (Mottram et al.~\cite{mott11}), which in turn implies
an inclination $\theta=\arcsin(\sqrt{\Mdyn/M_\ast})\simeq19\degr$.
An even smaller $\theta$ is needed to justify $M_\ast>24$~\Msun. While
an almost face on disk is consistent with the findings of Molinari
et al.~(\cite{mol98}), the SE--NW orientation of the disk (and
associated outflow) axis is not. In fact, the maps in Fig.~4 of
Molinari et al.~(\cite{mol98}), as well as the large-scale maps of Wu
et al.~(\cite{wu05}), suggest that the SiO flow is oriented NE--SW. In
addition, the luminosity of a $>$24~\Msun\ star should exceed the observed
luminosity of 3000~\Lsun. We thus believe that the model
consisting of Keplerian disk with expansion
is not acceptable.

\subsubsection{Rotating disk}

\begin{figure}
\centering
\resizebox{8.5cm}{!}{\includegraphics[angle=0]{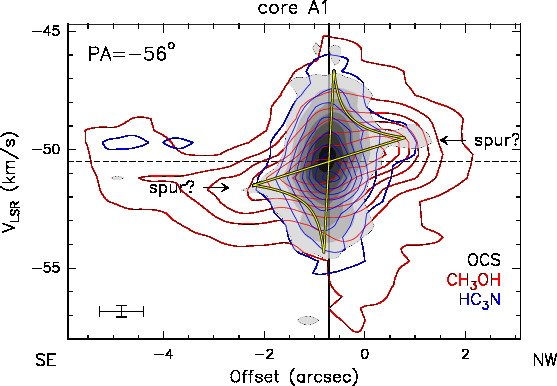}}
\caption{
Plot of the line intensity as a function of velocity and position along
the dashed line in Fig.~\ref{fvmcn} (position angle -56\degr). Different
colours correspond to different molecular species as indicated in the
bottom right. The cross in the bottom left gives the velocity and angular
resolution. The arrows indicate the possible ``spurs'' typical of the
pattern corresponding to Keplerian rotation. The yellow butterfly-shaped
curve is the region inside which emission is expected for
a Keplerian disk rotating about an 8~\Msun\ star. Contour levels range from
15 (5$\sigma$) to 123 in steps of 18~mJy/beam for OCS, from 17.5 (5$\sigma$)
to 332.5 in steps of 52.5~mJy/beam for \METH, and from 13.2 (4$\sigma$) to
151.8 in steps of 19.8~mJy/beam for \CYAC.
}
\label{fpva}
\end{figure}

An alternative scenario is that the velocity pattern in Fig.~\ref{fvmcn}
is due to the combination of a rotating disk oriented SE--NW, plus
a bipolar outflow in the NE--SW direction. The SE--NW velocity gradient
can be investigated through the position--velocity (PV) diagram along a
suitable cut (dashed line in Fig.~\ref{fvmcn}). Figure~\ref{fpva} shows
this diagram for the lines of three species and in all cases the plot is
vaguely reminiscent of the typical ``butterfly-shaped'' pattern consistent
with Keplerian rotation (see the yellow curves in the figure; e.g. Cesaroni
et al.~\cite{cesa05}).

Under the assumption of Keplerian rotation, the velocity and position of the
two putative ``spurs'' indicated in Fig.~\ref{fpva} can be used to estimate a
dynamical mass, which for a Keplerian disk is the mass of the central object.
We stress that we basically use the same approach as in Sect.~\ref{skepexp}
to derive an estimate of the stellar mass, with the only difference that
here the disk is oriented SE--NW (and hence responsible for the corresponding
velocity gradient), whereas in Sect.~\ref{skepexp} was oriented NE--SW.

For a rotation velocity of $\sim$1~\kms\
at a radius of $\sim$1\farcs5 or 0.036~pc, one obtains a stellar mass
of $\sim$8~\Msun\ (see the Keplerian pattern in Fig.~\ref{fpva}).
Although this value is in good agreement with the core luminosity of
3000~\Lsun\ estimated by Molinari et al.~(\cite{mol08b}), it is small
compared to the mass of the core (24~\Msun), inconsistent with the assumption of Keplerian
rotation, which requires the gas mass to be negligible with respect to
the stellar mass. This problem can be solved if the disk is sufficiently
inclined. Since $M_\ast = 8\,\Msun/\sin^2\theta$ (with $\theta$ angle
between the disk axis and the line of sight), the stellar mass can exceed
24~\Msun\ for $\theta<35\degr$. However, the luminosity of a $>$24~\Msun\
star is $>7\times10^4$~\Lsun\ (Mottram et al.~\cite{mott11}), much greater
than the estimated value of 3000~\Lsun.

All the above assumes that the disk is Keplerian. If one drops this
assumption and considers a self-gravitating disk, the dynamical mass
estimated above (8~\Msun) corresponds to the mass of the disk. As shown
above, for a suitable inclination this can match the mass of 24~\Msun\
obtained from the millimeter continuum. Since the disk is self-gravitating,
the stellar mass must be significantly less than the disk mass, consistent
with $\sim$9~\Msun\ derived from the luminosity of 3000~\Msun. In fact,
a self-gravitating disk oriented close to face-on is in agreement with
the findings of Molinari et al.~(\cite{mol98}). Also, the size appears
more consistent with those of ``toroids'' than to the typical diameters
($\sim$1000~au) of accretion disks around B-type stars (see Beltr\'an \&
de Wit~\cite{beldew}). Finally, a butterfly shaped PV diagram is not in
contradiction with a self-gravitating disk because its rotation curve
could mimic that of a Keplerian disk (e.g. Bertin \& Lodato~\cite{belo99},
Douglas et al.~\cite{doug13}, Ilee et al.~\cite{ile18}).  

It is worth stressing that the uncertainties on the estimate of the
dynamical mass are large. For example, assuming a radius of 2\arcsec\ and
a corresponding rotation velocity of 1.5~\kms, one obtains 25~\Msun. The
core mass is also uncertain, as it is inversely proportional to the
dust absorption coefficient (assumed 0.9~cm$^2$g$^{-1}$ by BEU18), which
might be underestimated by a factor $\sim$2 (see Table~1 of Ossenkopf \&
Henning~\cite{osshen}). Furthermore, the gas-to-dust mass ratio could be
100 instead of 150 (assumed by BEU18 in their calculations), which may cause
an overestimate of the mass by a factor 1.5.

Despite all these caveats, we believe that the hypothesis of a
self-gravitating disk oriented SE--NW and rotating about a $\sim$9~\Msun\
star is plausible. We remark that in this case, the NE--SW
orientation of the disk axis would be consistent with that of the large-scale
flow imaged by Wu et al.~(\cite{wu05}).

\subsubsection{Bipolar outflow}

A third possibility to explain the observed velocity gradient is that of
a compact bipolar outflow. Evidence for an outflow was provided also by
Molinari et al.~(\cite{mol98}) in terms of broad wings of the HCO$^+$(1--0)
and SiO(2--1) lines. Since no clear bipolar structure was identified,
these authors proposed that the outflow is directed close to the line of
sight. This is at odds with the small velocity difference ($\sim$1~\kms)
between the red- and the blue-shifted emission observed by us, because
typical flow speeds are much greater ($>$10~\kms) and projection effects should
not matter if the flow is directed close to the line of sight.
If an outflow is traced by the SiO emission, we believe that it should be
very compact and with a significant inclination with respect to the line of
sight.

To help distinguish between the disk and outflow scenarios we consider
Fig.~\ref{ftmcn}. Interestingly, the temperature rises from the center to
the border of the core along the direction of the SE--NW velocity gradient,
while the reverse occurs for the column density. This behaviour seems more
compatible with a bipolar outflow than a rotating disk, because in the
latter the temperature is expected to decrease with increasing radius.
As a matter of fact, it would not be surprising if \MCN\ was tracing an
outflow, as several studies have detected \MCN\ also in outflows (e.g. Leurini et
al.~\cite{leu11}, Busquet et al.~\cite{busq14}, Palau et al.~\cite{palau17}).

We can derive the parameters of the outflow, assuming that all the
material traced by the \MCN\ line participates in the expansion.  As a
first step, we estimate the abundance of \MCN, $X_{\rm CH_3CN}$, from the
ratio between the total number of \MCN\ molecules and the total number of
H$_2$ molecules. The latter is obtained from the mass of 24~\Msun\ from
Table~\ref{tparms}, while the former can be computed by integrating the
\MCN\ column density over the map in Fig.~\ref{ftmcn}.  We obtain $X_{\rm
CH_3CN}\simeq10^{-10}$, an order of magnitude less than the typical value of hot molecular
cores, but plausible for an object in an earlier evolutionary phase
(see Gerner et al.~\cite{gern14}).
Then we calculate the momentum of the flow from the expression
\begin{equation}
 P\,\cos i = \sum_j \frac{N_j}{X_{\rm CH_3CN}} (V_j-V_{\rm sys}) \, \mu \, m_{\rm H} \, d^2 \, \delta\Omega
\end{equation}
where $i$ is the inclination angle of the outflow axis with respect to
the line of sight, the sum is extended over all the pixels of the map in
Fig.~\ref{ftmcn}, $N_i$ and $V_i$ are the \MCN\ column density and velocity
at pixel $j$, $V_{\rm sys}$ the systemic velocity ($\sim$~--50.7~\kms),
$\mu=2.8$ the mean molecular weight, $m_{\rm H}$ the mass of the hydrogen atom,
$d$ the distance (4.9~kpc), and $\delta\Omega$ the solid angle of the
pixel. We obtain $P\,\cos i\simeq9$~\Msun~\kms. The corresponding timescale
of the flow, $t_{\rm out}$, is given by the ratio between the SE--NW size of
the region mapped in Fig.~\ref{ftmcn} and the maximum velocity range observed
over the same region: $t_{\rm out}\,\tan i = 1\farcs9/1.8~\kms = 0.045~{\rm
pc}/1.8~\kms = 2.4\times10^4$~yr. From this one obtains the momentum rate
$\dot{P} \, \cos^2i/\sin i = 3.7\times10^{-4}$~\Msun~\kms~yr$^{-1}$.
Assuming momentum conservation in the flow, this value should be valid
for the whole outflow, despite the fact that our estimate is obtained for
the small region traced by the \MCN\ emission.

Taken at face value and using the relationship between $\dot{P}$ and luminosity
determined by Maud et al.~(\cite{maud15}) for a distance-limited sample of outflows
from massive YSOs ($\Log[\dot{P}]=-4.8+0.61\,\Log[L]$),
the momentum rate implies a YSO luminosity of $\sim$175~\Lsun, far less
than the estimate of $3\times10^3$~\Lsun\ obtained by Molinari et al.~(\cite{mol08b}),
which instead corresponds to $\dot{P}=2.1\times10^{-3}$~\Msun~\kms~yr$^{-1}$.
To match this value, the outflow inclination must be $i\simeq70\degr$.
Such a large angle cannot be ruled out a priori, but is inconsistent with
the claim of Molinari et al.~(\cite{mol96}) that the outflow axis lies close
to the line of sight. Moreover, the SE--NW orientation of the
putative outflow in the plane of the sky is roughly perpendicular to that of the bipolar flow
imaged by Wu et al.~(\cite{wu05}) in the \CO(1--0) line on the arcmin scale.
While comparison between scales that differ by more than an order of magnitude
must be taken with caution, we believe that the outflow interpretation for
the observed \MCN\ velocity field is probably the least likely of all those
previously discussed by us.

Based on all the above,
we conclude that the \MCN\ emission in A1 is unlikely to trace an outflow
and we are thus inclined to prefer the disk scenario. It is also possible
that the observed velocity field is contributed by an accretion flow (through
the disk), as suggested by the red-shifted self-absorption seen in the
line profiles, as discussed later in Sect.~\ref{sclust}.

\subsection{Core A2}

Figures~\ref{fmoms} and~\ref{fvmcn} show that the LSR velocity increases
slightly from A1 to A2.
Figure~\ref{fmoms} also shows that the velocity dispersion
is significantly larger in A1 than in A2, while Figs.~\ref{flines}
and~\ref{foutf} demonstrate that the emission in all lines observed by us,
as well as in the continuum, is weaker in A2.
All of this supports our choice to distinguish this core from A1.

Another important feature denoting a difference between the two cores is the
chemical richness. From Table~\ref{tcores}, one sees that, while all listed
species are detected in A1, only 2 of them are clearly revealed in A2.
Finally, Fig.~\ref{ftmcn} demonstrates that both column density and
temperature decrease going from A1 to A2, while the latter is half as massive
as the former.

In conclusion, it seems unlikely that A2 is a starless core, given the
relatively large temperature (typically $\sim$60~K) obtained from \MCN.
However, the lower mass and turbulence suggest that any YSO in A2 could be less
massive and/or in an earlier evolutionary phase than in A1.

\subsection{Core B}
\label{scb}

This core is clearly separated from A1 only in the continuum and \METH\
emission. Although B is a fainter emitter in most tracers with respect
to A1 and A2, the mass and temperature are comparable to those of A1 and
A2, hinting at a substantial similarity between these cores. The most
striking feature of B is the ratio $M_{\rm H_2}/M_{\rm vir}$, which is
the only one above unity. We will comment on this fact in more detail in
Sect.~\ref{sclust}, but we can already draw the conclusion that B must be
virialized or on the edge of collapse and in all likelihood, this core is
bound to evolve into a status analogous to that of A1 and A2. Indeed, the
lack of emission in most of the observed lines (see Table~\ref{tcores})
appears consistent with the idea that this core is in an early evolutionary
phase.

\subsection{Core C}
\label{scc}

\begin{figure}
\centering
\resizebox{8.5cm}{!}{\includegraphics[angle=0]{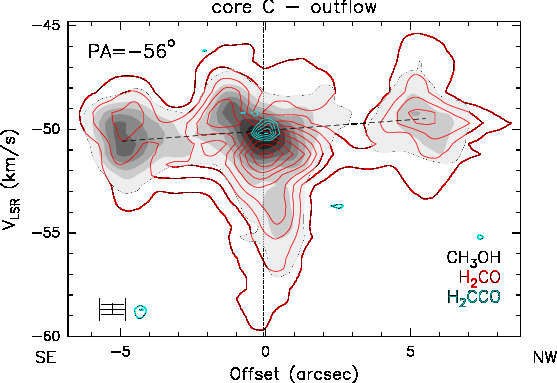}}
\caption{
Plot of the line intensity as a function of velocity and position along
the dashed line in Fig.~\ref{foutf} (position angle --56\degr). Different
colours correspond to different molecular species as indicated in the
bottom right. The cross in the bottom left gives the velocity and angular
resolution.
The dotted vertical line marks the position of core C, while the
dashed line outlines the velocity trend observed in the
\FORM\ and \METH\ lines. Contour levels range from 15 (5$\sigma$) to 225
in steps of 30~mJy/beam for \METH, from 18 (5$\sigma$) to 338 in steps of
40~mJy/beam for \FORM, and from 16 (4$\sigma$) to 32 in steps of 3~mJy/beam
for \KETE.
}
\label{fpvc}
\end{figure}

The most interesting feature of core~C is its possible association with the
bipolar outflow described in Sect.~\ref{sline}. In Fig.~\ref{fpvc} we show
the PV diagram of the emission along the dashed line in Fig.~\ref{foutf},
in three different lines. Several comments are in order for this plot. First
of all, there is a velocity trend, outlined by the dashed line in the
figure. Then it is worth noting that the \KETE\ emission tracing core C is
centrally located with respect to the SE and NW lobes (detected in \METH\
and \FORM) not only in space but also in velocity. Finally, the \METH\
and \FORM\ emission at the center is highly asymmetric in velocity,
with a broad blue wing, evident also from the spectrum in Fig.~\ref{fspts}.
At the same time, the lack of an equally broad red wing could be explained
by the presence of red-shifted self-absorption in a core undergoing infall.
This is consistent with the blue-shifted ``spot'' visible in
Fig.~\ref{fmoms}a, an indication of infall (see e.g. Mayen-Gijon et al.~\cite{magi14}).
These facts suggest that indeed the YSO powering the flow could be embedded
in an infalling core characterized by large velocity dispersion.
Whether
this core is indeed core~C (i.e. the one identified through the \KETE\
emission) is unclear, as the latter is slightly offset from both the outflow axis
(see Fig.~\ref{foutf}) and the blue-shifted ``spot''. In any case, we believe that this outflow cannot
be associated with a very massive object, due to the weakness of the line
emission from the lobes and the lack of a well defined core/YSO powering it.

\subsection{Cores D and E}

As one can see in Fig.~\ref{flines}, core D is only well defined in
the \METH\ map, while other tracers present, at most, a tail of emission
towards it. Similar to core D, also core E is clearly identified in only one
tracer, this time the continuum emission.
In summary, both cores have small masses and are recognizable
as such only in one tracer. However, there is an important difference
between the two: the velocity dispersion in D is much greater than that
in E, as witnessed by the FWHM of the \METH\ line, which is
1.8 times
larger
in D than in E (see Table~\ref{tparms}). We note that the line
FWHM in D
(3.6~\kms)
is even greater than that in A1
(3.2~\kms),
where various signposts indicate
star formation activity.

\section{Discussion}
\label{sdis}

\subsection{A cluster of massive stars in the making?}
\label{sclust}

Based on the analysis presented above, we consider the possibility that
\I\ consists of multiple star-forming cores in an early stage of
their evolution. Molinari et al.~(\cite{mol08b}) have provided evidence that
this object is indeed in a protostellar phase, prior to the formation of an
\HII\ region. However, their analysis assumes the presence of a single YSO,
which we associate with A1, because this core seems to be the most active
in terms of star formation and is hence likely to dominate the luminosity
and mass estimates used by Molinari et al.~(\cite{mol08b}) to establish
the evolutionary phase of \I.

We have shown that A1 is surrounded by other cores, most of which could
contain embedded YSOs undergoing accretion. The first question we are
asking is whether core formation proceeded through thermal
Jeans fragmentation. This can be checked by comparing the Jeans length
of the overall region with the mean separation of the
cores. The latter is $\Delta\simeq1\farcs2=5900$~au, whereas the former
can be computed from $\lambda_{\rm J}({\rm au})=1.16\times10^6 \sqrt{T/n_{\rm H_2}}$,
assuming the mean temperature ($T\simeq30$~K) from \METH\
and deriving the gas density ($n_{\rm H_2}\simeq2.3\times10^6$~\cmc) from the ratio
between the total core mass ($\sim$70~\Msun\ -- see Table~\ref{tparms})
and the volume occupied by the cores. The diameter of this volume (assuming
spherical symmetry) is equal to the size ($\sim$4\arcsec) of the region
over which the cores are distributed on the plane of the sky. We obtain
$\lambda_{\rm J}\simeq4200$~au. This value is only indicative, because
in the pristine cloud both the density and temperature were lower than those at
the present time. In consideration of these uncertainties, we believe that
$\lambda_{\rm J}$ is comparable to $\Delta$ and thermal Jeans fragmentation
could be a viable mechanism for the formation of the cores, consistent with
the results other studies of massive cores at similar spatial scales (Palau
et al.~\cite{palau15,palau18}; Ohashi~\cite{oha18}; Li et al.~\cite{li19}).

\begin{figure}
\centering
\resizebox{8.5cm}{!}{\includegraphics[angle=0]{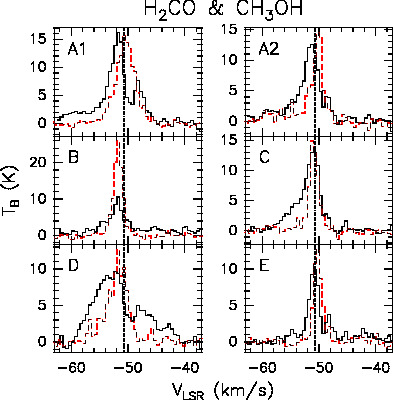}}
\caption{
Spectra of the \FORM($3_{03}$--$2_{02}$) (black solid histogram) and
\METH($4_2$--$3_1$)~E1 (red dashed histogram) lines towards the
positions of the 6 cores identified in \I. The vertical dotted line
is shown to ease the comparison between spectra of different cores.
}
\label{fspts}
\end{figure}

The next question is whether the cores are in equilibrium.
To address this issue, we consider the ratios $M_{\rm H_2}/M_{\rm vir}$
in Table~\ref{tparms}.
These are $<$1 in all cases except for core B, which seems to
indicate that only one core could be effectively forming stars.
To establish if this is indeed the case, we have searched for evidence
of infall onto the cores. In Fig.~\ref{fspts} we show a comparison
between the optically thick \FORM($3_{03}$--$2_{02}$) line and the thinner
\METH($4_2$--$3_1$)~E1 line, towards all the cores. We stress that the
\FORM\ spectra are obtained by merging the NOEMA and 30-m data to also recover
the extended emission. Therefore, the dips in the profiles cannot be
ascribed to missing flux filtered out by the interferometer.

The \FORM\ transition presents red-shifted self-absorption towards almost
all cores, except B. In fact, the \FORM\ profiles are skewed towards
blue-shifted velocities with respect to the \METH\ line, whose peak velocity
we take as a proxy for the systemic velocity of the core. This indicates
the presence of red-shifted self-absorption, a well-known indicator of
infall (see e.g. Mardones et al.~\cite{mardo} and references therein),
which in turn suggests that the cores could be forming stars in
the main accretion phase. This result is apparently inconsistent with the
fact that all cores except B have $M_{\rm H_2}<M_{\rm vir}$. We propose
that, instead, the presence of infalling material could exert an external
pressure on the cores, sufficient to stabilize them and even trigger their
collapse. This scenario could explain why some of the cores are already
forming stars, despite the low value of $M_{\rm H_2}/M_{\rm vir}$.

In practice, we can take into account the effect of external pressure
using Eq.~(8) of Field et al.~(\cite{field}), where the surface density
is expressed as $M_{\rm H_2}/(\pi\,R_{\rm c}^2)$. The virial mass is obtained
from this equation as
\begin{equation}
 M_{\rm vir} = \frac{5}{6} \frac{R_{\rm c}}{G}\,\frac{(\Delta V)^2}{8\ln2}
 \left(1+\sqrt{1-\frac{p_{\rm e}}{p_0}}\right)
\end{equation}
where $R_{\rm c}$ is the core radius, $\Delta V$ the line FWHM, $G$ the
gravitational constant, $p_{\rm e}$ the external pressure,
$p_0=\frac{15}{16\pi}\frac{(\Delta V)^4}{(8\ln2)^2\,G\,R_{\rm c}^2}$, and we
assume that the core has constant density (i.e. $\Gamma=3/5$ in the
notation of Field et al.~\cite{field} -- see also MacLaren et
al.~\cite{mclar}). If $p_{\rm e}>p_0$ no equilibrium configuration can
be attained and the core collapses.

The external pressure on the core is the sum of the thermal pressure
of the envelope plus the ram pressure of the infalling gas\footnote{
Our approach of including the ram pressure in the external pressure on the
core surface is not strictly correct in the hydrostatic scenario of Field
et al.~(\cite{field}). However, it has been shown by Ballesteros-Paredes
et al.~(\cite{bapa11}) that such an approach yields results comparable to
those obtained from the correct treatment of infall.
},
namely
$p_{\rm e}=n_{\rm H_2}\,(k\,T + \mu\,m_{\rm H}\,\varv_{\rm inf}^2)$,
with $k$ Boltzmann constant, $T$ gas temperature, and $\varv_{\rm inf}$
infall velocity. We assume $\varv_{\rm inf}\simeq2$~\kms\ from the
difference between the emission peak and absorption dip in the spectra of
Fig.~\ref{fspts}, while $n_{\rm H_2}\simeq2.3\times10^6$~\cmc\ and $T=30$~K
are estimated as previously explained.  From these and the parameters in
Table~\ref{tparms}, we find that $p_{\rm e}\simeq9.7\times10^{-7}$~dyn~\cmq\
and $p_0\simeq5.4\times10^{-8}$--$9.4\times10^{-7}$~dyn~\cmq. Therefore,
in all cases $p_{\rm e}>p_0$ and the cores appear to be bound to collapse.

Under the assumption that all cores are collapsing, we compute
the accretion rates, $\dot{M}_{\rm acc}$, and corresponding accretion
timescales, $t_{\rm acc}$, from the expressions
$\dot{M}_{\rm acc}=[3/(8\,\ln2)]^{3/2}\Delta V^3/G)$ and
$t_{\rm acc}=M_{\rm H_2}/\dot{M}_{\rm acc}$, where $G$ is the gravitational
constant. The values are listed in Table~\ref{tparms}. While $\dot{M}_{\rm acc}$
is consistent with the values for stars $>$10~\Msun\ (see Beltr\'an \&
de Wit~\cite{beldew}), $t_{\rm acc}$ seems quite short compared to the
typical timescale for the formation of a massive star ($\sim$$10^5$~yr)
and would imply that the cores are short lived or very young. However,
on the one hand not all the core material is necessarily accreting onto
a star, and on the other hand the cores can be loading fresh material
from the surrounding envelope, as indicated by the evidence for infall
discussed above. Our estimate of $t_{\rm acc}$ is hence to be considered
a lower limit.

\begin{figure}
\centering
\resizebox{8.5cm}{!}{\includegraphics[angle=0]{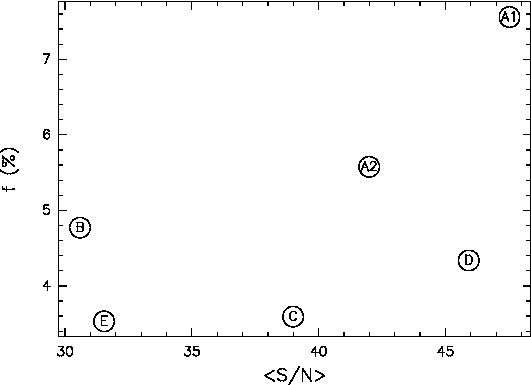}}
\caption{
Fraction of detected channels in the WIDEX spectrum of each core
versus the ratio between the mean intensity in the detected channels
and the noise of the corresponding spectrum. The names of the cores
are indicated inside the points.
}
\label{ffd}
\end{figure}

Another topic that is worth considering is the chemical richness
of the cores. A detailed analysis of the core chemical content requires
careful identification of the molecular lines detected in all cores. This
is a complex process that goes beyond the scope of our study. However, we
adopt a different approach, estimating the amount of molecular
line emission in the different cores. For each core we consider the spectrum
obtained with the broad-band correlator, WIDEX, and -- following Cesaroni et al.~(\cite{cesa17}) -- we calculate the fraction
of spectral channels, $f$, where emission is detected above 5$\sigma$.
In Fig.~\ref{ffd}, this parameter is plotted versus the corresponding
mean S/N, obtained as the ratio between the mean flux over the detected
channels and the 1$\sigma$ RMS of the spectrum. In this plot, for the
same value of S/N, sources with higher values of $N_{\rm det}/N_{\rm tot}$
are more line rich and, presumably, also more chemically rich. Therefore,
we can state that D is less rich than A1, and A2 is richer than C. As for
B and E, they appear line poor, but their S/N is worse than for the others
and it is thus possible that part of the line emission is not detected
only because it is weaker than in the other cores.

With all the above in mind, we can conclude that some chemical diversity
is present in the \I\ region. However, the spread in $f$ among the cores
is significantly less than that found by Cesaroni et al.~(\cite{cesa17})
in their sample of hot molecular cores (see their Table~3). While in their
case the values of $f$ are an order of magnitude greater, they span a wider
range (a factor 4.6) than in our sample (a factor 2.1). This yields a twofold
result: on the one hand, the chemistry in our cores is less evolved than
in typical hot molecular cores; on the other hand, our cores do not differ
much each other, which suggests that they might form the same type of stars.

In conclusion, we believe that our results provide convincing evidence
that the multiple cores in \I\ represent a
massive cluster\footnote{
With ``massive cluster'' here we mean a stellar cluster with
at least one massive (O-type or early B-type) YSO.}
in the
making. Although quite similar in mass (apart from C), the cores
appear to be in (slightly) different evolutionary stages, given their
differences in terms of temperature and velocity dispersion.

Is the purported cluster gravitationally bound? A rough estimate of its
stability can be obtained from the virial theorem. We assimilate
the core cluster to a bunch of mass particles, characterized
by the masses and line-of-sight velocities given in Table~\ref{tparms}.
From the standard deviation of the velocity distribution ($\sim$0.49~\kms)
and the maximum separation among the cores ($\sim$3\arcsec), one
obtains an equivalent virial mass of $\sim$10~\Msun, to be compared with the
total mass of the cores of $\sim$70~\Msun. Despite the large uncertainties
of the method, the difference between the two masses appears large enough
to indicate that the core cluster is in all likelihood gravitationally bound.

\subsection{Comparison with other high-mass star-forming regions}

The scenario depicted above is reminiscent of the case of W33A, investigated by Maud
et al.~(\cite{maud17}). Both sources, when observed at higher resolution,
reveal a cluster of cores, possibly in different evolutionary phases, and
do not present iron-clad evidence of Keplerian rotation, although the presence
of disks on smaller scales is suggested in W33A by detailed modelling
(Izquierdo et al.~\cite{izq18}). Also, in W33A accretion onto the cores
appears to proceed through a spiral-like filament, while in our case the line
profiles hint at infall onto most of the cores. Despite these similarities,
there are also differences between \I\ and W33A, the most important of which
are the spatial scales investigated in the two cases. The spatial resolution
is $\sim$10 times worse in \I\ ($\sim$4400~au) than in W33A ($\sim$500~au),
and the cores are distributed over a region of $\sim$3500~au in W33A as opposed to
$\sim$20000~au in \I. Moreover, W33A is $\sim$10 times more luminous but
$\sim$20 times less massive, and seems more chemically rich than \I. All
these features reinforce our conviction that \I\ is a young massive cluster
in the making, with a (proto)stellar content yet in an early evolutionary
phase (as predicted by Molinari et al.~\cite{mol98,mol08b}), whereas
W33A represents a more evolved stage.

More similar to our source is probably IRAS\,05358+3543, where at least 4
protostellar cores have been identified by Beuther et al.~(\cite{beu07}).
Although the spatial resolution is $\sim$900~au, better than
in our case, the core cluster spans a region of $\sim$16000~au, alike to \I.
The bolometric luminosity ($\sim$$6\times10^3$~\Lsun) and total mass of
all cores ($\sim$34~\Msun; see Table~3 of Beuther et al.~\cite{beu07})
are also similar to those of our source. Despite these similarities, the
two objects also present two substantial differences: no free-free
continuum emission has been detected from the cores in \I, whereas
a hypercompact \HII\ region is found in IRAS\,05358+3543; and multiple
outflows are known to be associated with the cores in this source,
unlike our case where only questionable evidence of two bipolar outflows
has been found. These differences seem to
suggest that IRAS\,05358+3543 is in a more advanced stage with respect to
\I. However, one should keep in mind that no deep, sub-arcsecond continuum
observation at centimeter wavelengths has been performed towards \I.
It is worth noting that the presence of the two \HII\ regions shown in
Fig.~\ref{fbig} could make it difficult to detect faint free-free continuum
emission from any putative hypercompact \HII\ region deeply embedded in
one of the cores.

\section{Summary and conclusions}
\label{scon}

In the context of the CORE project (P.I. Henrik Beuther) we have performed
observations of the continuum and molecular line emission at $\sim$1.4~mm from
the high-mass star-forming region \I. This is believed to be the youngest of
the whole CORE sample. We have revealed six cores and possibly an outflow from
a low-mass star. One of the cores (A1) could contain a self-gravitating disk
rotating about a $\sim$9~\Msun\ star, responsible for most of the 3000~\Lsun\
estimated for \I. The other cores are less massive and colder, but all of them
appear to be on the edge of collapse. The accretion rates estimated from the
velocity dispersions are consistent with those typical of high-mass star-forming
regions. We conclude that \I\ contains a sample of cores that are forming, or
are bound to form, a cluster of massive stars.

\begin{acknowledgements}
It is a pleasure to thank Daniele Galli for stimulating discussions
and Sergio Molinari for critically reading the manuscript.
We thank the IRAM technical staff for their support in this project.
DS acknowledges support by the Deutsche Forschungsgemeinschaft through SPP
1833: ``Building a Habitable Earth'' (SE 1962/6-1).
HB, AA, JCM, and SS acknowledge support from the European Research Council under the
Horizon 2020 Framework Program via the ERC Consolidator Grant CSF-648505.
RK acknowledges financial support via the Emmy Noether Research Group on
Accretion Flows and Feedback in Realistic Models of Massive Star Formation
funded by the German Research Foundation (DFG) under grant no. KU 2849/3-1
and KU 2849/3-2.
AP acknowledges financial support from UNAM-PAPIIT IN113119 grant, M\'exico.
RGM acknowledges support from UNAM-PAPIIT Programme IN104319.
This work is based on observations carried out under project number L14AB008
with the IRAM interferometer and 30-m telescope. IRAM is
supported by INSU/CNRS (France), MPG (Germany) and IGN (Spain).
XCLASS development is supported by BMBF/Verbundforschung through the Projects
ALMA-ARC 05A11PK3 and 05A14PK1 and through ESO Project 56787/14/60579/HNE.

\end{acknowledgements}

\end{document}